\documentclass[pra,aps,twocolumn,superscriptaddress,nofootinbib,nobalancelastpage]{revtex4}

\usepackage{graphicx}
\usepackage{amssymb}
\usepackage{amsmath}
\usepackage{color}

\newcommand{\ket}[1]{\left|{#1}\right\rangle}
\newcommand{\bra}[1]{\left\langle{#1}\right|}

\newcommand{\ketbra}[2]{\left|{#1}\right\rangle\left\langle{#2}\right|}
\newcommand{\trb}[1]{\textrm{tr}\left\{#1\right\}}

\newcommand{\beq}{\begin{equation}}
\newcommand{\eeq}{\end{equation}}
\newcommand{\eqn}[1]{\begin{eqnarray} #1\end{eqnarray}}
\newcommand{\ee}{&=&}
\newcommand{\nn}{\nonumber}

\begin{document}

\title{Measurement-Based Noiseless Linear Amplification for Quantum Communication}

\author{Helen M. Chrzanowski}
\affiliation{Centre for Quantum Computation and Communication Technology\\
Department of Quantum Science, Australian National University, ACT 0200, Australia}
\author{Nathan Walk}

\affiliation{
Centre for Quantum Computation and Communication Technology\\
 School of Mathematics and Physics, University of Queensland, St Lucia, Queensland 4072, Australia}
\author{Syed M. Assad}

\author{Jiri Janousek}
\author{Sara Hosseini}
\affiliation{Centre for Quantum Computation and Communication Technology\\
Department of Quantum Science, Australian National University, ACT 0200, Australia}
\author{Timothy C. Ralph}
\affiliation{
Centre for Quantum Computation and Communication Technology\\
 School of Mathematics and Physics, University of Queensland, St Lucia, Queensland 4072, Australia}
\author{Thomas Symul}
\author{Ping Koy Lam}
\affiliation{Centre for Quantum Computation and Communication Technology\\
Department of Quantum Science, Australian National University, ACT 0200, Australia}

\begin{abstract}
Entanglement distillation is an indispensable ingredient in extended quantum communication networks. Distillation protocols are necessarily non-deterministic and require advanced experimental techniques such as noiseless amplification. Recently it was shown that the benefits of noiseless amplification could be extracted by performing a post-selective filtering of the measurement record to improve the performance of quantum key distribution. We apply this protocol to entanglement degraded by transmission loss of up to the equivalent of 100km of optical fibre. We measure an effective entangled resource stronger than that achievable by even a maximally entangled resource passively transmitted through the same channel. We also provide a proof-of-principle demonstration of secret key extraction from an otherwise insecure regime. The measurement-based noiseless linear amplifier offers two advantages over its physical counterpart: ease of implementation and near optimal probability of success. It should provide an effective and versatile tool for a broad class of entanglement-based quantum communication protocols.
\end{abstract}
	
\maketitle

The impossibility of determining all properties of a system, as exemplified by Heisenberg's uncertainty principle \cite{Heisenberg:1927p8605} is a well known signature of quantum mechanics. It results in phase and amplitude fluctuations in the vacuum, enables applications such as quantum key distribution and is at the heart of fundamental results such as the no-cloning theorem \cite{Wootters:1982p8701}, quantum limited metrology \cite{Giovannetti:2011p3530}, and the unavoidable addition of noise during amplification \cite{Caves:1982p2405,Caves:2012p8233}. This last constraint means even an ideal quantum amplifier cannot be used for entanglement distillation \cite{Bennett:1996p8160,Horodecki:1997p8158,Browne:2003p153} which is a critical step in the creation of large scale quantum information networks \cite{Duan:2001p8161,Kimble:2008p5456}. 

Distillation protocols, originally conceived for discrete variables \cite{Bennett:1996p8160,Horodecki:1997p8158}, proved initially more elusive in the continuous variable (CV) regime. The most experimentally feasible, and theoretically well studied, class of CV states and operations are the Gaussian states and the operations that preserve their Gaussianity \cite{Weedbrook:2012p5160}. Protocols that distill Gaussian states were discovered \cite{Browne:2003p153,Eisert:2004p8182} involving an initial non-Gaussian operation that increases the entanglement followed by a `Gaussification' step that iteratively drives the output towards a Gaussian state. More recently noiseless linear amplification has been identified as a simpler method of distilling Gaussian entanglement \cite{proc-disc-2009,Ralph:2011p2764,Walk:2013p8604}.

The noiseless linear amplifier (NLA) avoids the unavoidable noise penalty by moving to a non-deterministic protocol.  This ingenious concept and a linear optics implementation have been proposed \cite{proc-disc-2009, Marek:2010p7535,Fiurasek:2009p7350} and experimentally realised for the case of amplifying coherent states \cite{Xiang:2010p1449,Ferreyrol:2010p2545,Ferreyrol:2011p2541,Zavatta:2010p4623}, qubits \cite{Osorio:2012p8806,Kocsis:2012p7641,Micuda:2012em}, and the concentration of phase information \cite{Usuga:2010p7394}. All of these were extremely challenging experiments, with only Ref.\cite{Xiang:2010p1449} demonstrating entanglement distillation and none directly showing an increase in Einstein-Podolsky-Rosen (EPR) correlations \cite{Reid:2009du}. Moreover the success probability of these experiments was substantially worse than the maximum set by theoretical bounds. In the context of quantum key distribution (QKD) Refs.\cite{Fiurasek:2012p7670,Walk:2013p403} proposed the possibility of implementing a non-deterministic measurement-based NLA (MB-NLA) to improve performance. This represents a significant advantage as the difficulty of sophisticated  physical operations can be moved from a hardware implementation, where one must suffer penalties related to source and detector efficiencies, to a software implementation where we are limited primarily by quantum theory and the statistics of our sample. Here we apply this protocol to EPR entanglement and observe improvements in the measured correlations consistent with distillation of the entanglement. We emphasise that this method is only equivalent to entanglement distillation for certain applications. Specifically, it is only situations where the desired distillation operation immediately precedes the measurement of the target mode that the two are indistinguishable. 

We first derive some general conditions on the limits to implementing arbitrary quantum operations on an ensemble by conditionally filtering the measurement results. Using this method we experimentally implement an MB-NLA protocol achieving significant distillation with a much improved probability of success. Furthermore we illustrate the critical benefit of distillation in combating decoherence by considering the distribution of EPR entanglement through a lossy channel. We measure an output level of entanglement that  exceeds the maximum achievable without distillation, even if one could use a perfect initial entangled state.

In any quantum information application the final result is always some classical measurement record, drawn from a set of possible outcomes $\{k\}$ and described by some probability distribution $p(k)$. In an application where the proposed distillation would take place immediately prior to measurement, for example in quantum key distribution, one could imagine emulating the operation on an ensemble via post-selective measurements. We first consider the process of emulating arbitrary operations via conditioning on measurements in a general setting before describing the results of Refs.~\cite{Fiurasek:2012p7670,Walk:2013p403} in which an explicit procedure applicable to the NLA was proposed. Our analysis will allow us to clarify some of the previous work as well as showing that the $g^{\hat{n}}$ operator key to the operation of the NLA, is particularly well suited to emulation via post-selection.

Consider an arbitrary quantum map applied to an incoming state $\rho$ which can be written using the Kraus decomposition as \cite{Hellwig:1970p8209}.
\eqn{\mathcal{E}(\rho) = \sum_iE_i \rho E^\dag_i}
where $\{E_i\}$ are the Kraus operators.
Note that this decomposition is valid for any completely positive operator including those which do not preserve the trace, as is the case with many useful conditional operations in quantum optics including photon addition and subtraction and the NLA. In the latter case the Kraus operators fail to satisfy the usual relation $\sum_i E_i^\dag E_i = \mathbb{I}$, with the extra information needed to restore conservation of probability being the success probability, $P$, of the conditional process \cite{Ferreyrol:2012p8191}.

If this map is immediately followed by a positive-operator valued measure (POVM) described by operators $\{\pi_k\}$ the corresponding probability distribution is,
\eqn{
p(k) \ee \trb{\pi_k \sum_i E_i\rho E_i^\dag} \nn \\
\ee \trb{\tilde{\pi}_k\hspace{1mm} \rho \label{ps}} 
}
where $\tilde{\pi}_k=\sum_i  E_i^\dag \pi_k E_i =\tilde{\mathcal{E}}(\pi_k)$ are a new set of POVM elements obtained by applying the mapping $\tilde{\mathcal{E}}$ to the desired output POVM set. What Eq.\ref{ps} tells us is that we may obtain the statistics of a POVM set $\{\pi_k\}$ upon an arbitrarily transformed state $\mathcal{E}(\rho)$ by conditioning upon measurements made with a transformed set $\{\tilde\pi_k \}$ on the original input state.

Although the above procedure is quite general, it is not arbitrary in that it does not allow the reconstruction of any desired POVM set in combination with any desired operation. If one wishes to implement arbitrary operations $\mathcal{E}$, one is restricted to certain final POVM sets $\{\pi_k\}$ and vice versa. Intuitively we expect that in order to correctly reconstruct the statistics of an arbitrary POVM upon an arbitrary state it is necessary to obtain maximum information about that state, i.e. to make measurements capable of complete tomographic reconstruction. This requirement can be derived by considering Eq.~(\ref{ps}) and noting that the POVM set with which one must actually measure, $\{\tilde{\mathcal{E}}(\pi_k)\}$, is not necessarily physical for arbitrary $\mathcal{E}$ and $\{\pi_k\}$. The unphysicality occurs because some of the operations that we wish to emulate are not themselves physical. For example the NLA itself, as will be discussed later, is trace increasing \cite{supp}. If one demands access to arbitrary operations, then a sufficient condition on $\{\tilde{\mathcal{E}}(\pi_k)\}$ would be that it maps to physical output states and is capable of uniquely determining an arbitrary CP map. This is precisely the same condition required of a POVM set for it to be classified as informationally complete (IC) \cite{Prugovecki:1977p7335,Busch:1989p7336}. Conversely, if one is only able to experimentally realise a certain POVM set then one is limited in the range of operations that can be faithfully implemented. Further details regarding the status of IC-POVM's as a sufficient condition for implementation via post-selection in combination with non-deterministic operations are given in the supplementary information \cite{supp}.

Noiseless amplification is commonly defined as the ability to increase the amplitude of an unknown coherent state without any noise penalty, effecting the transformation $\ket{\alpha}\rightarrow \ket{g\alpha}$ with $g>1$. By considering  the annihilation and creation operators describing a bosonic mode it becomes clear that such a transformation would violate the     canonical commutation relations $[\hat{a},\hat{a}^\dag] = 1$. Consistency with quantum mechanics can be restored if one instead implements a non-deterministic version of this transformation $\ket{\alpha}\bra{\alpha} \rightarrow P\ketbra{g\alpha}{g\alpha} + (1-P)\ketbra{0}{0}$ which succeeds with probability $P$. Provided the success is heralded, one may enjoy the benefits of entirely noiseless amplification at least some fraction of the time. 

As was shown in Refs.~\cite{proc-disc-2009,Fiurasek:2009p7350} just such a transformation is performed by the operator $g^{\hat{n}}$ where $\hat{n}=\hat{a}^\dag \hat{a}$ is the number operator. Furthermore, acted upon one arm of a two-mode gaussian EPR state written in the number basis as
\beq
\ket{\chi,\chi} = \sqrt{1-\chi^2}\sum_n \chi^n\ket{n}\ket{n}
\eeq
where  $\chi \in \{0,1\}$ characterises the entanglement, we find the operation results in
$g^{\hat{n}} \ket{\chi,\chi}\bra{\chi\chi} \rightarrow P \ketbra{g\chi,g\chi}{g\chi,g\chi}+ (1-P)\ketbra{0,0}{0,0}$. In other words the entanglement is probabilistically increased. Applying the NLA to an EPR state that has been distributed through a lossy channel (Figure \ref{figure1}(a)) results in an output with a greater degree of initial entanglement that appears to have suffered less loss \cite{proc-disc-2009}. We note that although $g^{\hat{n}}$ appears Gaussian in the sense of being quadratic in annihilation/creation operators it is in fact non-unitary and unbounded. These properties are also the reason that such an operation falls beyond the purview of the no-go theorem \cite{Eisert:2002p466,Fiurasek:2002p467,Giedke:2002p468} which states that Gaussian entanglement cannot be distilled via purely Gaussian operations. In fact an exact implementation of $g^{\hat{n}}$ would necessitate a success probability of zero. However, when considering a given set of input states one may explicitly construct physical operations which have arbitrarily high fidelity with $g^{\hat{n}}$ while succeeding with a finite probability. The most intuitive version of this method, proposed in Ref.~\cite{proc-disc-2009} and utilised in subsequent experiments, is to use a generalised quantum scissors scheme \cite{Pegg:1998p1507} and truncate in the photon-number basis faithfully amplifying low energy input states that have negligible higher order terms. However these truncated experiments are by no means trivial, with all demonstration limited to the single photon case except for \cite{Kocsis:2012p7641} in which two stages were achieved. 
Thus it would be extremely valuable to devise an easier method of implementing the distillation, albeit for a more restricted set of applications. Here we implement a measurement-based version of this protocol (Figure \ref{figure1}(b)) where the original state is first measured using heterodyne detection upon Bob's side.  Then a sub-ensemble is post-selected according to a filter function defined by the desired NLA gain. 
\begin{figure}[htbp]
\begin{center}
\includegraphics[width = 8cm]{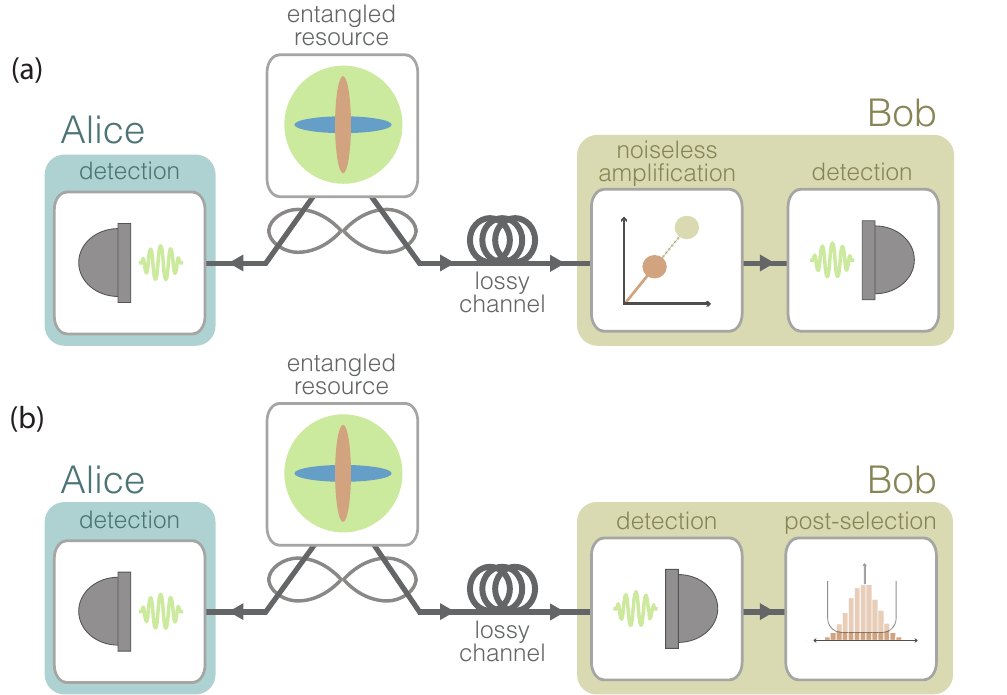}
\caption{ Equivalent methods of entanglement distillation with (a) physical and (b) measurement-based noiseless linear amplifiers. Two-mode EPR entanglement is represented by two  orthogonally juxtaposed squeezed state. One arm of the EPR entanglement is transmitted through a lossy channel before being noiselessly amplified. In the physical implementation (a) a quantum scissor setup is used to implement the probabilistic amplification before the final measurement. In the measurement-based implementation (b) a post-selective filter is used to keep a remaining fraction of data. 
\label{figure1}} 
\end{center}
\end{figure}

The MB-NLA is also in compliance with the no-go theorem of Refs.~\cite{Eisert:2002p466,Fiurasek:2002p467} in that we do not distill free propagating Gaussian entanglement. Nevertheless it is remarkable that, in certain circumstances, we achieve useful results utilising only hardware from the experimentally friendly Gaussian toolbox. We emphasize that, with respect to the post-selected ensemble, the protocol works shot-by-shot.

The exact filter function corresponding to $g^{\hat{n}}$ can be derived following Ref.\cite{Fiurasek:2012p7670} by considering the coherent state projection, or the Q function, on an arbitrary input state $\rho$,
$Q_\rho(\alpha) = \bra{\alpha}\rho\ket{\alpha}/\pi$.
Recalling that the action of the NLA on a coherent state is given by \cite{proc-disc-2009},
\eqn{g^{\hat{n}}\ket{\alpha} = e^{\frac{1}{2}(g^2-1)|\alpha|^2}\ket{g\alpha}}
we can write down the Q function of the amplified state $\rho'$,
\eqn{Q_{\rho'}(\alpha)\ee \bra{\alpha}g^{\hat{n}}\rho g^{\hat{n}}\ket{\alpha} \nn \\
\ee e^{(g^2-1)|\alpha|^2} \bra{g\alpha}\rho\ket{g\alpha} \nn\\
\ee e^{(1-1/g^2)|\beta|^2} \bra{\beta}\rho\ket{\beta}.\label{het}}
where $\beta = g\alpha$. This equation allows us to determine the particular probabilistic filter and rescaling we must apply to the original heterodyne data in order to obtain the same output as a heterodyne measurement applied to the same input state after noiselessly amplification with a gain $g$. Clearly for $g>1$ the filter defined above does not qualify as a sensible weighting probability as it is always greater than 1. Thus we must renormalise to some cut-off thereby implementing an approximation to the ideal  operation. This is analogous to the fact that although the success probability for $g^{\hat{n}}$ has to be zero, one can experimentally achieve a good approximation of an ideal NLA with finite probability. In the measurement-based picture, this corresponds to implementing an approximate operation while keeping a finite fraction of the data after post-selection. In both cases, however, the approximation can be made arbitrarily close to perfect whilst retaining a finite success probability. 

\begin{figure}[htbp]
\begin{center}
\includegraphics[width = 8.5cm]{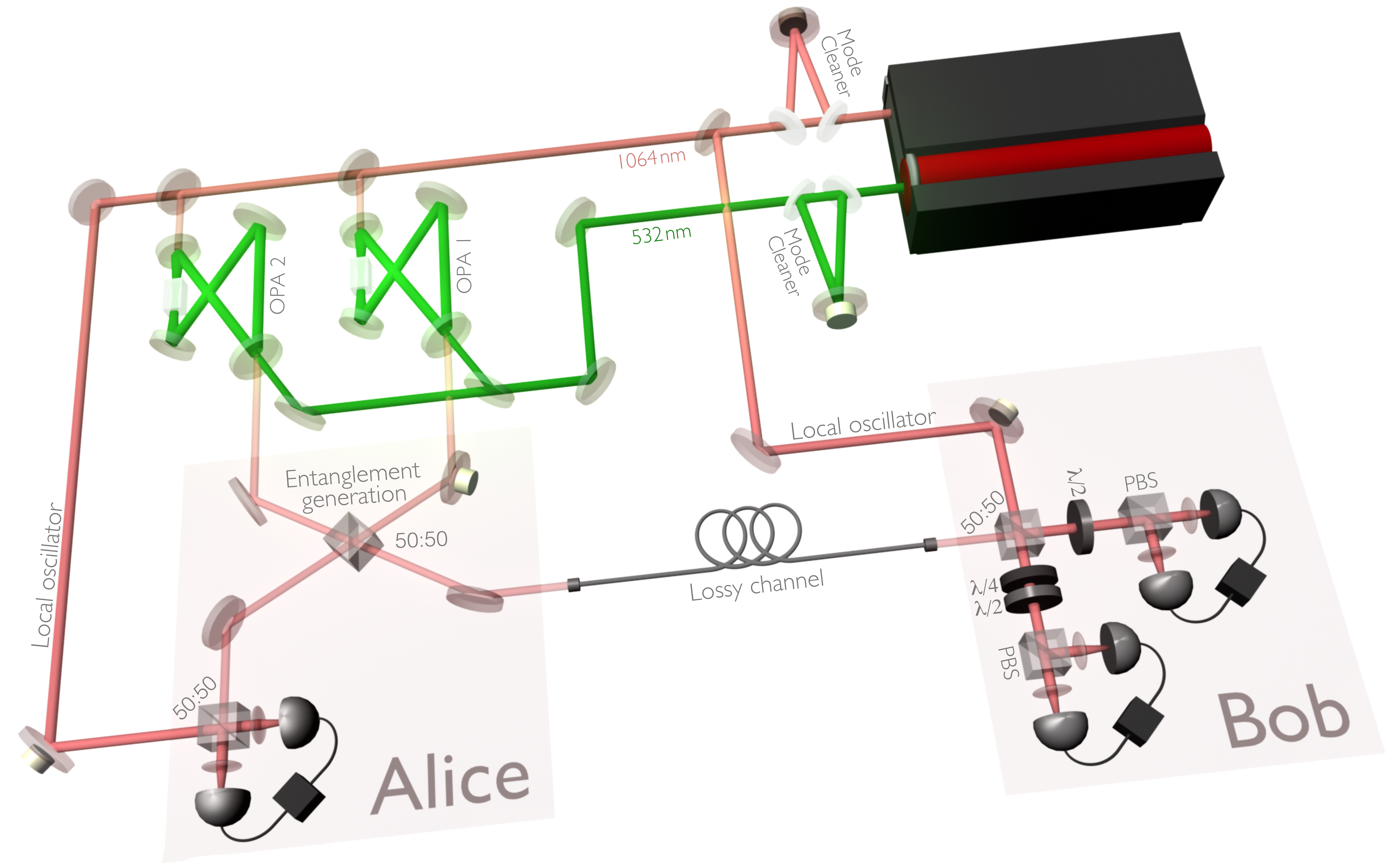}
\caption{Experimental setup of the measurement-based NLA. A laser provides both 1064nm and 532nm fields. These fields are spatial and frequency filtered to the quantum noise limit for the sideband detection frequencies between the range of $3 - 4$~MHz. The 1064nm field is used as the seed and local oscillator fields for two identical degenerate bow-tie optical parametric amplifiers (OPAs), whilst the 532nm light is used as the pump field. Two amplitude squeezed states are produced and combined on a $50 \! : \! 50$ beam-splitter. With their relative phase locked in quadrature, the beam-splitter produces two-mode EPR state at the output. The entangled beams are sent to Alice locally and through a transmission channel to Bob remotely. Alice performs homodyne detection of her optical states, alternating between conjugate quadratures. Bob on the other hand, performs a heterodyne detection of his state, simultaneously measuring both conjugate quadratures. A sample size of $10^9$ data points is used for all quadrature measurements. \label{figure2}}
\end{center}
\end{figure}

\begin{figure*}[htbp]
\begin{center}
\includegraphics[width = 17cm]{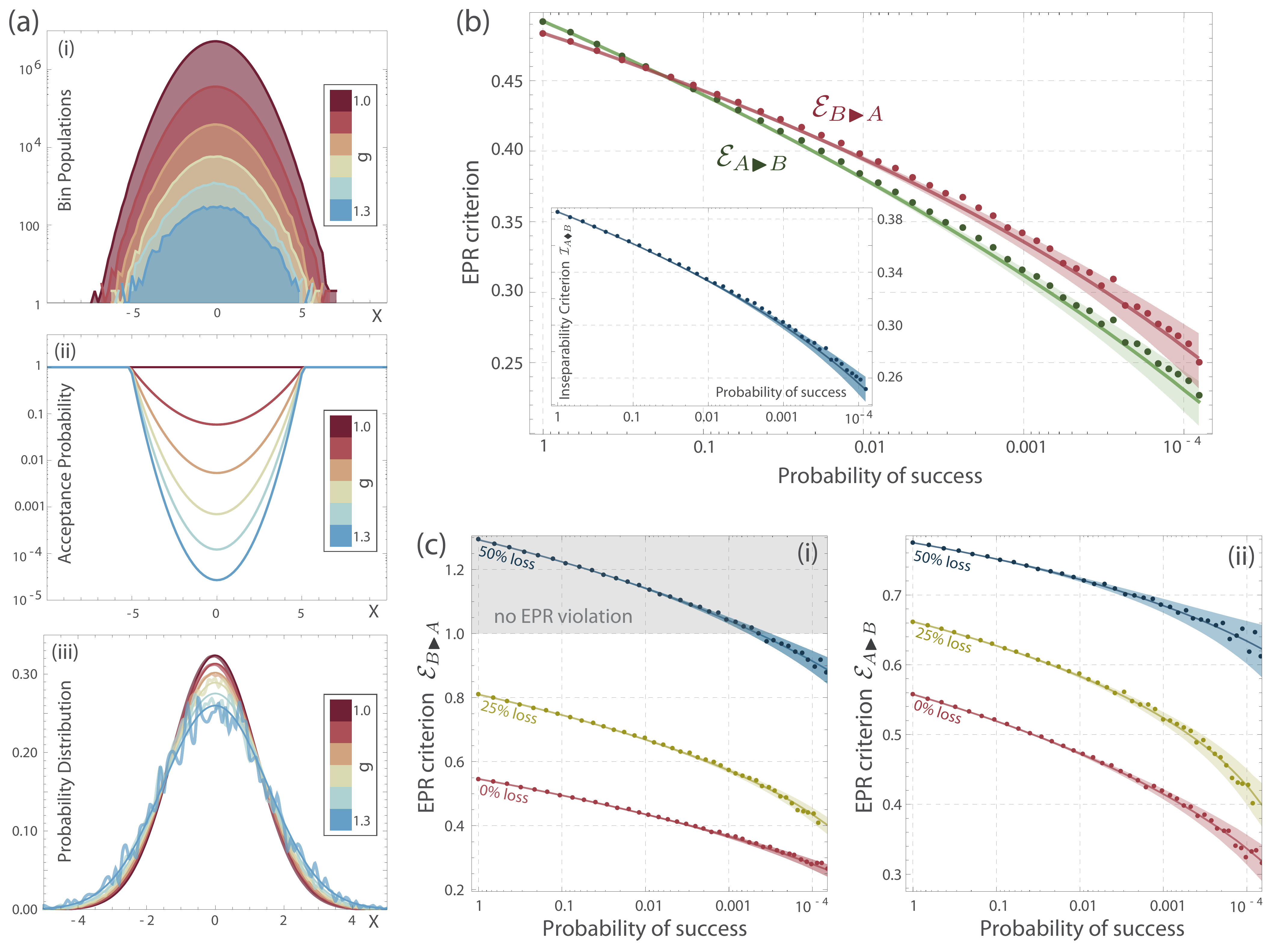}
\caption{{\bf (a)} Measurement-based NLA performed on the receiver, Bob's, experimental data: (i) Acceptance probability function of the post-selective filters used to obtained (ii) the resulting measurement histograms, and (iii) the final normalised probability distributions. The gain, $g$, is increased by selecting filter function with increasingly lower acceptance probability. As the gain is increased, the variance of Bob's final measurement distribution increases. This corresponds to a stronger two-mode EPR entanglement. {\bf (b)} EPR criterion as a function of post-selection success probability for the direct ($\mathcal{E}_{B \blacktriangleright A} $, red) and reverse ($\mathcal{E}_{B \blacktriangleright A}$, green) inferences for input state with an initial EPR entanglement strengths of  $0.484 \pm 0.001$ and  $0.492\pm 0.001$ respectively. Data points presented are the post-selected ensemble average of 10 experimental runs.  The solid lines shows the theoretical distillation of an ideal implementation of $g^{\hat{n}}$ given the same input state. Shading represents a $2\sigma$ confidence interval on the variance of the implemented filter. Inset shows the effect of the distillation on the inseparability criterion, $\mathcal{I}_{A \blacklozenge B}$, with the same data set. {\bf (c)} Effect of EPR entanglement distillation as a function of probability of success for different losses ($0\%$, $25\%$ and $50\%$). \label{figure3}} 
\end{center}
\end{figure*}

\begin{figure}[htbp]
\begin{center}
\includegraphics[width = 8.5cm]{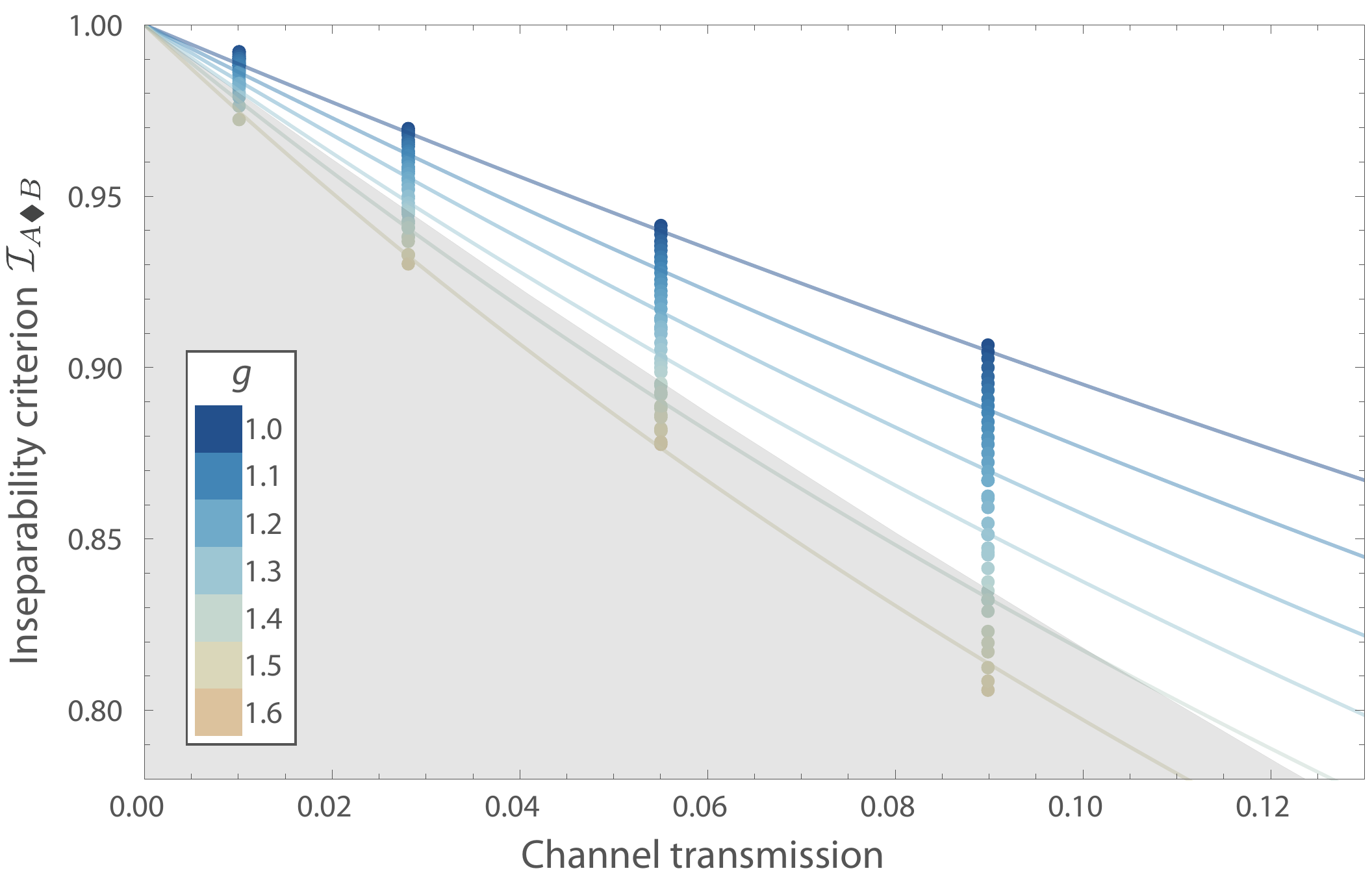}
\caption{Improvement in the inseparability criterion of the two-mode EPR state for a series of lossy channels. For each transmissivity, a series of post-selection corresponding to an NLA gain (specified by the legend) are applied. The boundary of the shaded area describes the theoretical inseparability of a perfect EPR state - infinitely squeezed - subject to the same channel transmissivity. Post-selection allows access to an entangled resource beyond that accessible with even perfect initial resource. The solid lines represent theoretical inseparability of our input state, with an applied post-selection filter of a defined gain ($g= 1, 1.1,\ldots 1.5$) as a function of the channel transmission.\label{figure4}}
\end{center}
\end{figure}

The filter function, or acceptance probability, of the $g^{\hat{n}}$ modified post-selection filter with a finite cutoff is given by,

\begin{equation}
P(\alpha) = \left\{ \begin{array}{cc}
e^{\frac{1}{2} (|\alpha|^2 - |\alpha_{C}|^2)(1-g^{-2})},& \alpha <\alpha_C\\
1,& \alpha \geq \alpha_C \label{psfunction}
 \end{array} \right .
\end{equation}
where $\alpha = \tfrac{1}{\sqrt{2}}(x + ip)$ is the coherent state projection for each heterodyne measurement. Examples of the function $P(\alpha)$, and the resultant histograms and probability distributions for various gains are shown in Fig. 3(a). As the fidelity of the truncated filter with the ideal $g^{\hat{n}}$ operation is input state dependent, we choose a finite cutoff $\alpha_C$ that optimises post-selection rates whilst preserving high fidelity with the ideal filter. This also ensures that output distributions remain statistically close to a normal distribution, allowing us to only consider the second order moments of the measured distribution to characterise the correlations \cite{supp}. For each input state we implement a cut-off of between $4-5$ standard deviations of Bob's measured state. Whilst the energy of Bob's input state defines a theoretical maximum gain of our amplifier \cite{Walk:2013p8604}, the size of the initial data ensemble effectively places a stronger restriction on the maximum gain that can be applied whilst retaining reasonable statistical precision. This is analogous to the rapidly diminishing success probability of the physical NLA for larger amplifier gains \cite{Xiang:2010p1449} though less severe.

A similar effect is observed in \cite{Pandey:2013wb} which explores the ultimate limits of noiseless amplifiers that involve a cut-off in photon number. The probabilistic nature of any implementation of $g^{\hat{n}}$ must be carefully accounted for when considering metrological applications, as the noiseless amplification must be balanced against the reduced sample size of the measurement \cite{Pandey:2013wb}. For example, even the success probabilities achieved here would render the operation unsuitable for metrology protocols that scale with the square root of the number of measurements.

Let us consider that Alice and Bob share a symmetric two-mode EPR state, with no transmission loss between either of their measurement stations. In this scenario, a MB-NLA on Bob's side is indistinguishable from an implementation on Alice's side, and the observed distillation should be symmetric for both parties. A two-mode EPR state demonstrates Einstein-Poldosky-Rosen type correlations if it demonstrates a Reid EPR criterion $\mathcal{E} < 1$ which is a condition on the product of the conditional quadrature variances \cite{Reid:1989vm}. As such it is an inherently directional quantity and we borrow terminology from QKD, and refer to Bob's ability to infer Alice's state as the {\it direct} inference ($\mathcal{E}_{B \blacktriangleright A} = V_{x_A|x_B}V_{p_A|p_B}$), and the converse as the {\it reverse} inference ($\mathcal{E}_{A \blacktriangleright B}= V_{x_B|x_A}V_{p_B|p_A}$). Either $\mathcal{E}_{A \blacktriangleright B}<1$ or $\mathcal{E}_{A \blacktriangleright B}<1$ is a sufficient but not necessary condition for entanglement. We also examine the symmetric inseparability criterion of Duan {\it et. al.} \cite{Duan:2000ts} denoted $\mathcal{I}_{A \blacklozenge B}$. For Gaussian states $\mathcal{I}_{A \blacklozenge B}<1$ is both a necessary and sufficient condition for entanglement.

We experimentally prepare a two-mode EPR resource by interfering two identical amplitude squeezed-states on a 50:50 beam splitter with their relative phase controlled to be in quadrature. The two arms of the resulting two-mode EPR state are then sent to two independent measurement stations that we identify with Alice and Bob (see Fig.\ref{figure2}). We implement a MB-NLA on Bob's subsystem, amounting to a heterodyne measurement of the amplitude and phase quadratures $X$ and $Y$, followed by the appropriate post-selection procedure. Alice implements a homodyne measurement of her subsystem, alternating between measurement of $X$ and $Y$. These measurements allow characterisation of the covariance matrix of the two-mode EPR state, and thereafter, the corresponding EPR criterion \cite{Reid:1988wh} $\mathcal{E}_{A\blacktriangleright B}, \mathcal{E}_{B\blacktriangleright A}$ and inseparability criterion \cite{Duan:2000ts}, $\mathcal{I}_{A \blacklozenge B}$. 

Our initial entangled resource demonstrates an EPR criterion violation of $\mathcal{E}_{A \blacktriangleright B} = 0.484 \pm 0.001$ and  $\mathcal{E}_{B \blacktriangleright A} = 0.492 \pm 0.001 $ with an initial ensemble size of $10^{9}$ data points. We then apply the post-selection function of Eq.~(\ref{psfunction}) followed with a rescaling by $1/g$, emulating $g^{\hat{n}}$. A linear increase in the amplifier gain sees an exponential reduction in the probability of success but results in a more correlated subset of the measurement record, equivalent to a more entangled two-mode EPR state. Figure \ref{figure3}{\rm (b)} demonstrates our improvement in the EPR criterion as a function of the success probability. The declining probability of success as we apply increasingly larger gains to obtain stronger correlations manifests in increased statistical uncertainty. The solid line indicates the behaviour of an ideal implementation of $g^{\hat{n}}$ with the same input state \cite{Walk:2013p8604}, and the shaded area gives a $2 \sigma$ confidence interval on the theoretical EPR violation. For a post-selection probability of $8\!\times \!10^{-5}$ we obtain EPR criterions of $\mathcal{E}_{A \blacktriangleright B} =  0.25\pm 0.02 $ and  $\mathcal{E}_{B \blacktriangleright A} =  0.23 \pm 0.02 $. The asymmetry in the EPR criterions for the direct (green) and reverse (red) inferences arises due to slight variations in the purity of the two-subsystems; Bob's heterodyne measurement introducing additional loss. Figure \ref{figure3}{\rm (b)} also plots the inseparability criterion as a function of the success probability. We find excellent agreement between theory and experiment.

We also investigate the performance of the MB-NLA for noiseless amplification after a lossy channel. We experimentally introduce loss on Bob's subsystem, allowing us to model several lossy channels. Bob subsequently implements a MB-NLA. Figure \ref{figure3}{\rm (c)} demonstrates the performance of the post-selective NLA for a two-mode EPR state with moderate-loss on one subsystem. Figure \ref{figure3}{\rm (c) i.} plots the EPR criterion for the direct inference as a function of post-selection probability for a series of channel transmissions, whilst the reverse inference is plotted in Figure \ref{figure3}{\rm (c) ii}. For $25\%$ loss on Bob's channel, we find the post-selection allows us to fully compensate for any loss incurred, and measure EPR correlations beyond those of the original state. For moderate loss, and sufficient post-selection, we find Bob can obtain the same EPR violation as Alice, despite the asymmetry of their initial subsystems. For $50\%$ loss post-selection is shown to recover an EPR violation from an initial state that does not display an EPR violation.

Perhaps the most interesting regime for the performance of the MB-NLA occurs at very high loss. In Figure \ref{figure4} we plot the inseparability criterion of the two-mode EPR state as a function of the channel transmission encountered by Bob's subsystem. For each channel, we consider distillation using the MB-NLA with the maximum loss of 99\% equivalent to 100km of optical fibre \cite{fnote}. The boundary of the shaded area describes the theoretical inseparability of a perfect EPR state in the limit of infinite squeezing, subject to the same channel transmissivity. We find that post-selection allows access to  final EPR correlation that, without our protocol, are inaccessible even considering a perfect initial EPR resource.

\begin{figure}[htbp]
\begin{center}
\includegraphics[width = 8.5cm]{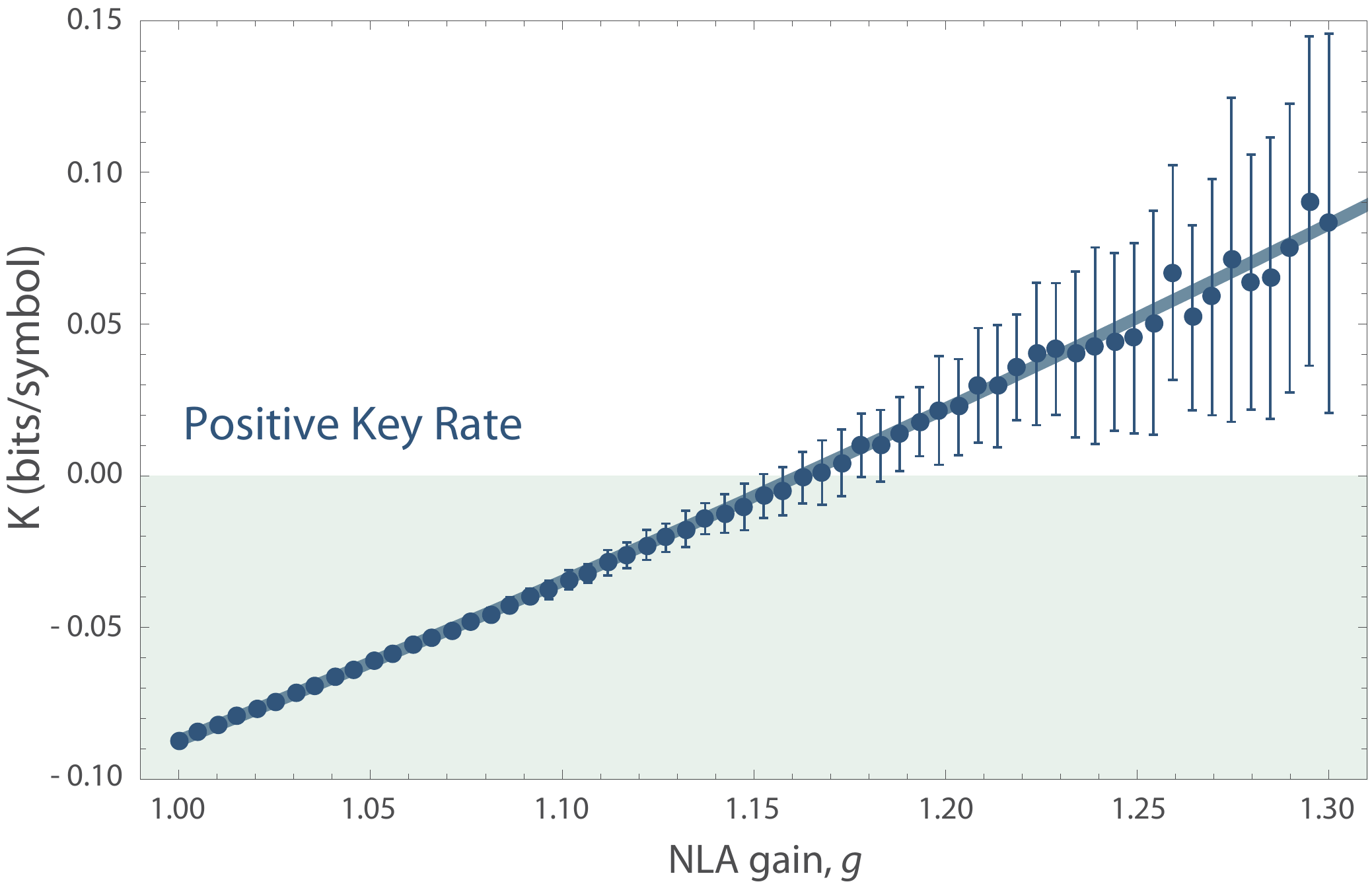}
\caption{Secret key rate as a function of the gain for a direct reconciliation CV-QKD protocol where both parties use heterodyne detection. The application of the MB-NLA allows the recovery of secure key distribution from an initially insecure situation. The error bars represent a $1\sigma$ confidence interval on the key rate.
\label{figure5}}
\end{center}
\end{figure}

Finally we turn to the application that sparked interest in this protocol, and investigate the performance of entanglement-based CV-QKD protocol including an MB-NLA \cite{supp}. We conduct a very cautious analysis in which all measured imperfections are attributed to the eavesdropper, such that our EPR source is interpreted as a pure EPR source followed by a decohering channel. In Fig. \ref{figure5}, we show that, starting from a situation in which key distribution is impossible, application of a sufficiently high gain restores security. Further discussions and a calculation of the key rate are given in \cite{supp}.

The primary significance of these results is two-fold. Firstly, we experimentally demonstrated the equivalence of the MB-NLA to the implementation of a physical NLA for entanglement distillation when considering scenarios where amplification is directly proceeded by measurement. This equivalence ensures this technique has immediate applications for CV-QKD, where the advantage of an NLA has already been studied \cite{Blandino:2012p5681,Fiurasek:2012p7670,Walk:2013p403}. Furthermore, we provided a generalised theoretical explanation of the conditions in which an arbitrary quantum operation could in principle be implemented upon an ensemble via post-selective measurements. Secondly, this equivalence is of great practical importance as the MB-NLA is significantly less demanding than the existing physical implementations of $g^{\hat{n}}$, where inefficiencies in sources and measurement restrict the physical NLA to very small input states\cite{proc-disc-2009,Xiang:2010p1449,Ferreyrol:2010p2545,Kocsis:2012p7641}. In contrast, the MB-NLA is more flexible in that it can be used on a wide variety of input states without experimental reconfiguration, and more scalable, in that by circumventing several experimental inefficiencies it achieves near optimal success probability for an implementation of $g^{\hat{n}}$ of arbitrary precision. Whilst there are clear restrictions on the scenarios where this MB-NLA can be substituted for its physical counterpart, when applicable, it is certainly advantageous to do so. The achievable entanglement distillation is now chiefly limited by the amount of data collected. Here, for feasible sample sizes, we demonstrate distillation of correlations in excellent agreement with close to the theoretical ideal performance of $g^{\hat{n}}$. For moderate loss channels we showed the recovery of EPR correlations from an entangled state, and applied to high loss channels demonstrated levels of entanglement that are impossible without a distillation process.

Many avenues for further research remain. Beyond the aforementioned applications in CV-QKD, the NLA could find use in other quantum communication protocols including teleportation and remote state preparation. This would be of particular interest as it would enable us to extend these conditioning distillation techniques to improve the quality of a still propagating, albeit unentangled, quantum mode. Furthermore our theory is sufficiently general to allow extensions to other conditional processes. For example using precisely the same setup described here it is also possible to implement the photon addition operation which has been extensively studied \cite{Fiurasek:2009p7350,Barbieri:2010p8804,Zavatta:2004p8805}. As well as targeting other operations one could also use this formalism to consider conditioning on different POVM sets, opening up many promising candidates for future applications. 

{\it Acknowledgements--}
We thank S. Rahimi-Keshari and A.P. Lund for helpful discussions. The research is supported by the Australian Research Council (ARC) under the Centre of Excellence for Quantum Computation and Communication Technology (CE110001027). P.K.L. is an ARC Future Fellow.

Author contributions
N.W. and T.C.R. developed the theory. H.M.C., S.M.A., J.J., S.H., T.S. and P.K.L. conceived and conducted the experiment. H.M.C., S.M.A. and N.W. analysed the data. H.M.C., N.W. and S.M.A. drafted the initial manuscript. P.K.L., T.C.R. and T.S. planned and supervised the entire project. All authors discussed the results and commented on the manuscript.

Additional information
First author statement for H.M.C. and N.W.: These two authors contributed equally to this work
Correspondence and requests for materials should be addressed to P.K.L. e-mail:Ping.Lam@anu.edu.au

Competing financial interests
The authors declare no competing financial interests.

\appendix

\section{Importance of informational completeness}
\label{ICPOVM}
By considering Eq. 2 in the main text we were able to deduce a sufficient condition for the successful implementation of an arbitrary operation via post-selection, namely that the measurement upon which we post-select be informationally complete (IC). A concrete example of this is to consider the two most readily available CV measurements, homodyne and heterodyne detection. While heterodyne detection is always IC, homodyne detection when restricted to a single quadrature (as opposed to Wigner function reconstruction where all quadratures are scanned in principle) is not. 

This clarifies the somewhat different results found in Refs. ~\cite{Fiurasek:2012p7670,Walk:2013p403}. While both works suggest using a Gaussian post-selection in order to virtually implement the NLA operation Ref.\cite{Fiurasek:2012p7670} conditions using IC heterodyne measurements, whereas in Ref.\cite{Walk:2013p403} the conditioning is done via homodyne measurements, randomly switching between two conjugate quadratures. The latter measurements are not IC and the corresponding post-selection protocol actually implements a noisier version of the NLA. Although this is still useful for key distribution because the noise is confined within a secure station for the purpose of demonstrating entanglement distillation we wish to remain as close as possible to the ideal operation. Thus applying the protocol of Ref.\cite{Fiurasek:2012p7670} is the superior choice for our task. 

Mathematically, the problem is apparent as soon as one attempts to derive an appropriate filter function in analogy to Eq.4 of the main text by considering a homodyne measurement upon the amplified state, $\rho'$,
\eqn{P(x)_{\rho'} = \bra{x}g^{\hat{n}}\rho g^{\hat{n}}\ket{x}.}
The action of the NLA on squeezed states has been calculated \cite{Gagatsos:2012p5682,Walk:2013p8604} where it was shown to displace and further squeeze the state. However, if one acts it upon the homodyne projectors, technically infinitely squeezed states, the resultant state is not a displaced $x$ eigenstate but rather, divergent and unphysical. This is the sense in which the the procedure described in the main text cannot always be applied to arbitrary measurements if the operations to be emulated are not themselves strictly physical. Further discussion of this point can be found in Ref.\cite{Walk:2013p8604}. 

More intuitively, we can see the extra noise as being due to our incomplete knowledge of the target state. Recalling once more,
\eqn{g^{\hat{n}}\ket{\alpha} = e^{\frac{1}{2}(g^2-1)|\alpha|^2}\ket{g\alpha}}
it becomes clear that the pre-factor induced by $g^{\hat{n}}$, which we are essentially mimicking via post-selection, depends at each shot upon both the $\hat{x}$ and $\hat{p}$ displacements. Only measuring one quadrature at a time makes it impossible to know this quantity, and hence the correct filter function cannot be determined. Attempting to filter based upon our partial knowledge results in a bigger state, but with the missing information inducing extra noise.


\section{Post-selection procedure} 

As previously discussed, $g^{\hat{n}}$ is unbounded for $g>1$, and therefore, whether physically or virtually, it cannot be implemented exactly. However, the virtual implementation, $g^{\hat{n}}$ can be emulated to arbitrary precision by truncating our original post-selection filter at an appropriate amplitude, $\alpha _C$. The resulting modified post-selection filter is given by
\begin{equation}
P(\alpha) = \left\{ \begin{array}{cc}
e^{\frac{1}{2} (|\alpha|^2 - |\alpha_{C}|^2)(1-g^{-2})},& \alpha <\alpha_C\\
1,& \alpha \geq \alpha_C
 \end{array} \right . \label{psfunction}
\end{equation}
where, $\alpha$ is obtained from the measured heterodyne outcome via $\alpha = \tfrac{1}{\sqrt{2}}(x + ip)$. Any measurement outcomes falling beyond the cutoff amplitude, $\alpha_C$, are kept with unit probability. Simply, Bob's measurement-based implementation of $g^{n}$ amounts to him obtaining a measurement outcome, $\alpha_i$ (from his heterodyne measurement), which he keeps or rejects with a probability specified by Eq. \ref{psfunction}. If his outcome $\alpha_i$ falls beyond the cutoff $\alpha_C$, he always keeps it. If we consider a two-mode scenario (virtual distillation), Bob also informs Alice to either keep or reject her state - measured or unmeasured.

Some care needs to be taken with the choice of $\alpha _C$ to ensure that the truncated approximation emulates $g^{\hat{n}}$ with high fidelity. Perfectly indistinguishable fidelity with the $g^{\hat{n}}$ can always be obtained by pushing $\alpha_C$ out to the largest measurement outcome encountered in the measurement ensemble \cite{Fiurasek:2012p7670}. This approach, however, scales very poorly with increasing ensemble size \cite{Fiurasek:2012p7670}.  We find that a compromise between excellent emulation of $g^{\hat{n}}$ and the post-selection probability can be achieved by implementing an $\alpha_C$ sufficiently large that the purity ($\frac{1}{\sqrt{\mathrm{det(\sigma)}}}$) of the post-selected state decreases consistent with an ideal implementation of $g^{\hat{n}}$. As an ideal implementation of $g^{n}$ results in a Gaussian mapping of the input state, by here ensuring that our emulation is consistent (within statistical error) any non-Gaussianity is necessarily small. This can be quickly verified by examining the size of the third (skewness) and fourth (kurtosis) order moments provides a quick verification of the normality of the distribution after post-selection. Figure \ref{figure1} gives the calculated skewness and kurtosis of the post-selected distribution (the results presented in Figure 3 (b) of the main text) as a function of post-selection probability. Owing to the symmetry of the filter it should not introduce skewness, but implementation of the filter without a sufficiently large cutoff may introduce a ``peakedness'' that might be quantified via the fourth moments. We also considered a Jarque-Bera test of the distribution to verify its normality to a confidence interval of 95\%. 

\begin{figure}[htbp]
\includegraphics[width = 8cm]{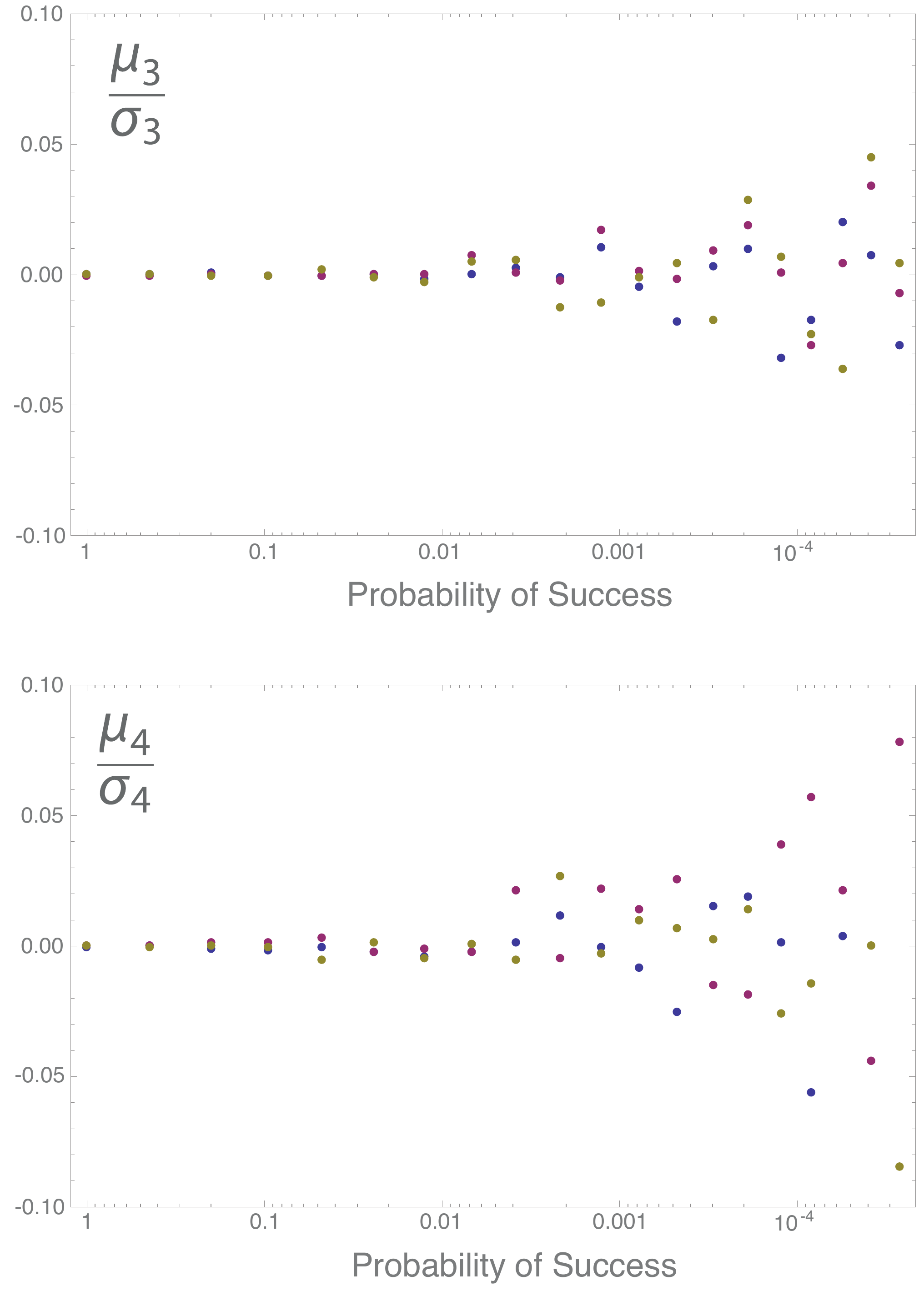}
\caption{ Third (skewness) and fourth (kurtosis) of the post-selected ensembles as a function of the probability of success. As the probability of success declines the uncertainty associated with estimating each moment increases, but there is no appreciable trend.} 
\label{figure1}
\end{figure}
It is also important to note that here, for a given input state, all emulations of $g^{\hat{n}}$ for varying $g$ use the same cut-off $\alpha_C$. We restrict ourselves to an $\alpha_C$ sufficiently big to accommodate the largest gain applied. One could potentially improve the post-selection rates by considering a gain dependence in the choice of $\alpha_C$.

\section{Computation of secret key rate}
The security proofs for protocols using Gaussian post-selection for arbitrary attacks in the asymptotic limit of large key lengths are given in \cite{Fiurasek:2012p7670,Walk:2013p403}. They essentially emerge from the fact that the previous results leading to asymptotic security proofs - the minimising of the key rate under collective attacks by assuming a Gaussian state for any measured CM \cite{Navascues:2006p805,GarciaPatron:2006p381} and the optimality of collective attacks in the asymptotic limit \cite{Renner:2009p1} - both hold even if the CVQKD protocol is not perfectly Gaussian. It is only that the bounds would become very pessimistic were the post-selection to exhibit a strong non-Gaussianity, but as explained above this is not the case here.

This level of analysis is the same as that employed in several CVQKD experiments \cite{Fossier:2009p385,Lodewyck:2007p374}, including the only previous demonstration of CV-QKD using entangled states \cite{Madsen:2012p6851}, but does not include all of the finite-size effects \cite{Leverrier:2010p150} and reconciliation and privacy amplification processes of a state of the art CV-QKD demonstration such as \cite{Jouguet:2013p8197}. Nevertheless it is sufficient to demonstrate the benefit of the MB-NLA for key distribution.

There are some subtleties involved with using genuine entangled states, as opposed to most CV-QKD experiments where a theoretical equivalence is established between a prepare and measure scheme (P\&M) and one involving real entanglement \cite{Grosshans:2003p526}. P\&M schemes can generally prepare states of extremely high purity, and the appropriate theoretical entangled state is of much higher purity than is currently experimentally feasible. This can be mitigated by an additional step where the decoherence in the source production is characterised and the purification of that noise is not attributed to the eavesdropper. This method was successfully employed by \cite{Madsen:2012p6851} who showed that the EPR based scheme actually showed improved robustness to channel noise in comparison to coherent state P\&M protocols. Completing the proof to include this step would be beyond the scope of this work, however we can still demonstrate the value of the MB-NLA while attributing all observed impurities to the eavesdropper. This means that our measured covariance matrices are interpreted as coming from a pure EPR source that has been transmitted through a lossy channel with thermal noise. The effective channel has a relatively low loss but high additional noise. In this situation the optimal protocol would be DR with heterodyne detection on both Alice and Bob's side. 

The secret key rate for this protocol is given by \cite{Navascues:2006p805,GarciaPatron:2006p381},
\eqn{K^{\blacktriangleright} = \beta I(A:B) - S(A:E)}
where $I(A:B)$ is the classical mutual information between quadrature measurements made by Alice and Bob, $S(A:E)$ is the Holevo quantity between Eve and Alice and $\beta \in [0,1]$ is the reconciliation protocol efficiency. Here we choose a value of $\beta = .98$, consistent with \cite{Madsen:2012p6851}. Given that the post-selected measurements are still very Gaussian, Alice and Bob's can be calculated using the formula,
\eqn{I(A:B) = \log\frac{V_A+1}{V_{A|B}+1}}
where $V_A$ is the measured homodyne variance on Alice's side and $V_{A|B}$ is the conditional variance of Alice's measurement given Bob's heterodyne detection. Eve's mutual information is given by \cite{Navascues:2006p805,GarciaPatron:2006p381},
\eqn{S(A:E) = S(AB) - S(B|A)}
where $S(AB)$ and $S(B|A)$ are the von Neumann entropies of the inferred state $\rho_{AB}$ and the conditional state following a heterodyne detection by Alice. In general, the von Neumann entropies are bounded by those of a Gaussian state with the same covariance matrix (CM). The Gaussian von Neumann entropy can be calculated straightforwardly from the symplectic eigenvalues of the CM \cite{Weedbrook:2012p5160}, and thus our key rate can be determined directly from the unconditional and conditional CM's or our system. The secret key rate is plotted as a function of the NLA gain if Fig. 5 in the main paper, demonstrating a transition from and insecure to a secure regime. As noted in \cite{Walk:2013p403,Blandino:2012p5681} the NLA acts to improve the transmission of the effective channel while actually increasing the noise, but in such a way as to create an information advantage between Alice and Bob. As explained in \cite{Walk:2013p8604} this can be seen as the amplifier distilling both the Alice-Eve and the Alice-Bob entanglement.


\begin{thebibliography}{46}
\bibitem[{fno()}]{fnote}
\bibinfo{note}{This figure is derived assuming a loss of .02dB per km of
  optical fibre leading to the relationship between transmission $T$ and
  distance $d$ of $T=10^{-.02 d}.}$
\end{thebibliography}
\end{document}